\documentclass[letterpaper,10pt]{article} 

\usepackage[utf8]{inputenc}
\usepackage{osameet3} 

\usepackage{amsmath,amssymb}
\usepackage[colorlinks=true,bookmarks=false,citecolor=blue,urlcolor=blue]{hyperref} 

\usepackage[draft]{fixme}

\begin{document}

\title{Symbiotic joint operation of quantum and classical coherent communications}

\author{Raphaël Aymeric, Yves Jaouën, Cédric Ware and Romain Alléaume}
\address{LTCI, T\'el\'ecom Paris, Institut Polytechnique de Paris, 19 place Marguerite Perey, 91120 Palaiseau, France}
\email{raphael.aymeric@telecom-paris.fr}
\copyrightyear{2022}

\begin{abstract}
We report successful joint operation of quantum and classical communications with shared hardware. Leveraging information learned from the classical DSP, low-noise quantum communications (0.009 SNU at 15 km) compatible with 15 Mbit/s QKD is demonstrated.
\end{abstract}

\section{Introduction} 
As Quantum Key Distribution (QKD) is pushing towards commercial maturity, the trade-off between system performance and cost becomes of central concern. Since deployment of dedicated dark fiber links for QKD is extremely costly, it is crucial to design QKD protocols and engineer QKD systems so that they can be deployed over existing telecom infrastructure with a limited cost overhead. While discrete-variable (DV) QKD enables key establishment over larger distances \cite{diamanti2016}, continuous-variable (CV) quantum communications is an appealing technology in which coherent receivers with high common mode rejection ratio, act as built-in filters. This allows to realize low-noise quantum communications (compatible with QKD at positive key rate) while being multiplexed with classical communication channels over a single fiber link \cite{kumar2015,eriksson2019}.
 
 Modern implementations of CV-QKD are now converging with classical coherent communications system designs, where high-speed receivers combined with DSP allow to digitally phase lock a local oscillator (LO) placed at receiver side, with the emitter laser. Such local LO (LLO) designs remove security loopholes that existed in previous CV-QKD designs, where the LO was transmitted \cite{marie2017self}. Phase reference sharing however becomes a challenge. Most recent LLO CV-QKD demonstrations  rely on reference signals (also known as pilot tones) multiplexed in time \cite{marie2017self,roumestan2021} or frequency \cite{kleis2017,laudenbach2019} with the quantum channel in order to precisely estimate carrier phase and frequency.

Moving towards tighter QKD integration over existing networks, it is interesting to determine whether a pure reference signal is necessary or if a classical information-carrying channel can provide solid estimates for the quantum channel without inducing excess noise beyond the null key rate threshold. In this work,we demonstrate for the first time a joint transmission over 15 km of a Quadrature Phase Shift Keying (QPSK) modulated classical channel and of a quantum channel carrying QPSK modulated QKD states for which carrier phase and frequency is directly recovered from the classical channel. Our quantum and classical signals share hardware at emission while we use dedicated low-noise, low-bandwidth detectors at reception for the quantum channel.  
 
\section{Joint transmission}

\subsection{Experimental setup}

\begin{figure}[t]
    \centering
    \includegraphics[width=\linewidth]{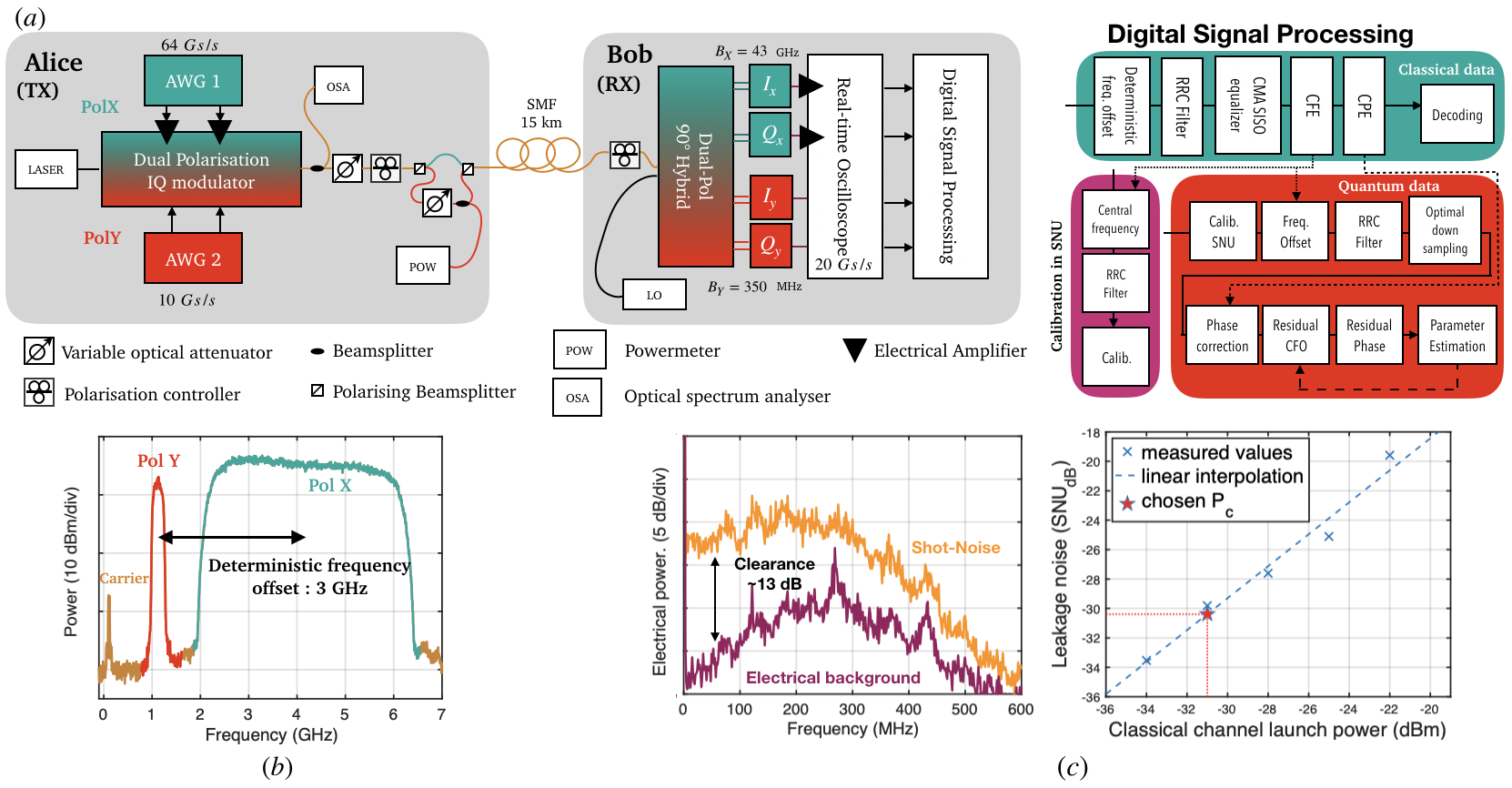}
    \caption{(a) Experimental setup and DSP. Classical channel is depicted in green and quantum channel in red. Calibration data is in purple. (b) Spectrum acquired on the OSA. (c) Shot-noise versus electronic noise clearance (left) and excess noise due to classical channel leakage (right). The leakage noise is estimated by computing the excess noise variance over a block, when the quantum signal is turned off. }
    \label{fig1}
\end{figure}

At Alice a low linewidth laser (100 Hz) source generates continuous-wave light. Two synchronised arbitrary waveform generators (AWGs) encode QPSK modulations at rates 4 GBaud and 250 MBaud and are respectively linked to the analog input of the X- and Y- polarisation of the dual polarisation IQ modulator. A root-raised cosine (RRC) filter with roll-off factor 0.2 and 0.1 for the quantum and classical channel is applied for spectral shaping of the signal. In order to reduce cross-talk the signals are digitally frequency shifted by $1$ and $4$ GHz respectively, this way there is no spectral overlap. This can be seen in Fig. \ref{fig1} (b). The classical channel power is tuned using a first variable optical attenuator (VOA) and set to launch power $P_{c} = -31$ dBm. The choice of $P_{c}$ is made as to generate minimal leakage noise on the quantum data (see Fig. \ref{fig1}(c)) while maintaining reliable communications on the classical channel. Then, a polarisation controller (PC) and a polarising beam splitter (PBS) are used to separate both polarisation tributaries. On the path corresponding to the quantum channel a second VOA is placed for tuning the quantum channel power before recombining both polarisations. The modulation variance of the quantum channel was taken as $V_{A} = 0.49$ SNU since positive key rates require low $V_{A}$ for QPSK modulation \cite{Lin2019,Denys2021explicitasymptotic}. The multiplexed signals are transmitted over a single-mode fiber spool of 15 km.

At the receiver, a polarisation controller aligns the signal polarisation with the dual polarisation phase diversity hybrid where the signal is mixed with a low linewidth (100 Hz) local oscillator (LO). The LO wavelength is tuned such that it is close to the quantum signal as thermal noise in the balanced photoreceivers increase at higher frequencies. The quantum channel is routed to the dedicated low noise photoreceivers with a clearance of approximately $\sim$ 13 dB (see Fig. \ref{fig1}(c)) while the classical signals are detected on standard high bandwidth (43 GHz) balanced receivers. The data is retrieved from the oscilloscope running at 20 Gs/s by blocks of 10 Megasamples corresponding to $1.25\times 10^{5}$ quantum symbols and $2\times 10^{6}$ classical symbols per block. The oscilloscope digitizes the analog data with an effective number of bits (ENOB) of $\sim$6, which can generate relatively high quantization noise for the quantum channel. 

\subsection{Digital signal processing}

The DSP routine processes the acquired data in the following order: classical data, shot-noise calibration data and quantum data (highlighted in green, purple and red respectively in Fig. \ref{fig1}(a)). The classical data processing begins by correcting the deterministic frequency offset with the quantum channel of 3 GHz. We then apply the matched filter. A CMA-based adaptive equalization operated in single-input single-output (SISO) is applied to mitigate inter-symbol interferences and downsamples the input signal from 5 to 1 sample/symbol. Then a classical Viterbi \& Viterbi 4\textsuperscript{th}-power algorithm provides the carrier frequency estimation (CFE) and phase estimation (CPE). Finally the QPSK symbols are decoded using hard decision.

 Shot noise calibration data is performed by turning off the quantum signal. This calibration is interleaved with quantum data acquisition, allowing to estimate shot noise almost in real time, following a technique implemented in \cite{kumar2015}.

\begin{figure}
    \centering
    \includegraphics[width=\linewidth]{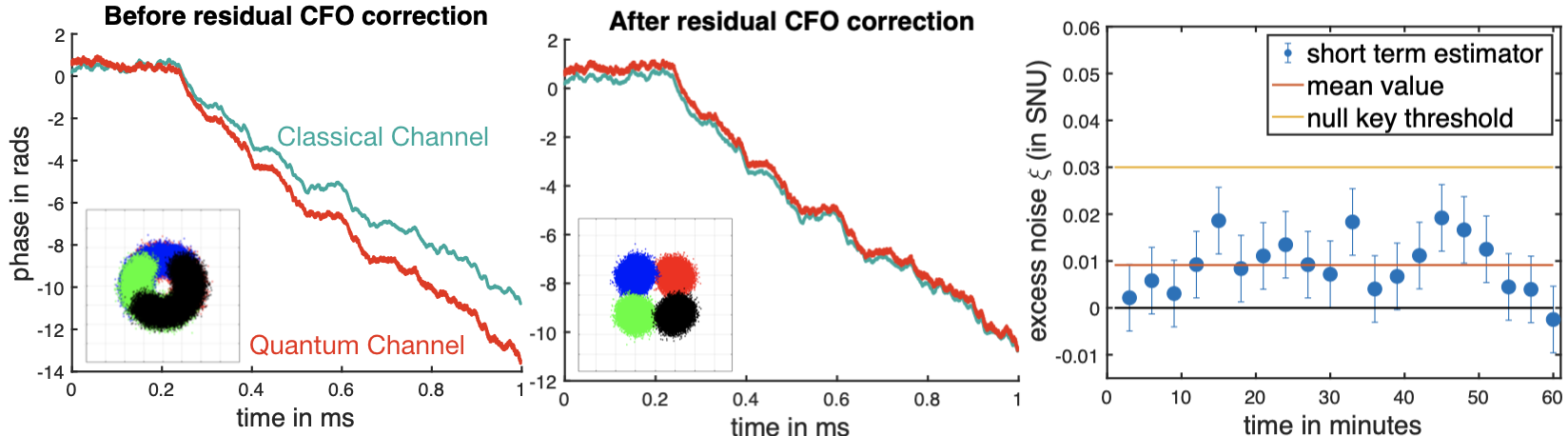}
    \caption{Left and middle plots show the residual frequency offset (455 Hz) on the quantum channel before and after correction. Right plot shows the excess noise ($\xi$) estimators over 3 blocks (i.e. $4.75\times 10^5$ symbols), with 3 sigma confidence intervals. Over one hour, the $\xi$ estimators have a mean value of $\xi_{B} = 0.009$ SNU and are well below the null key threshold (yellow line)}
    \label{fig2}
\end{figure}

 Finally, the quantum data is processed offline during the digital signal processing (DSP) step. First, we rescale Bob's quadratures in shot-noise units and correct the frequency offset. Then the matched filter is applied to the samples and the quantum signal is downsampled to 1 sample/symbol. The choice of the optimal symbol is determined relative to the classical data, thus the classical channel also provides synchronisation information the quantum channel. 
 Note that unlike in \cite{roumestan2021,laudenbach2019} no dedicated pilot tones are used. Next, because of clock time jitter of the DACs and ADCs, the 3 GHz spacing between quantum and classical channels fluctuates which leads to a residual frequency offset on the quantum data. We illustrate this in figure \ref{fig2} when the quantum channel is operated in a high intensity regime in order to estimate the phase directly on the quantum data. 
We solve this by exploiting the symbols revealed during parameter estimation, allowing Bob to perform an incremental frequency offset correction until the excess noise is minimal. This method provides good estimators of the residual frequency offset as can be seen in the figure. A similar method is employed to correct residual phase offset of the quantum constellation, which is shifted until the covariance between Alice and Bob's data is maximal. 

\section{Results and perspectives}

We conducted the experiment over one hour and recorded the BER on the classical data as well as the excess noise on the quantum data quadratures measured by Bob. Our results give an average excess noise at Bob's over all measurements of $\xi_{B} = 0.009$ SNU for $V_{A} = 0.49$ while the classical BER is measured to be less than $4\times 10^{-5}$, i.e. below FEC threshold. To simulate the use of off-the-shelf (100 kHz linewidth) lasers, we also processed our data with artificially added phase, which yielded a negligible excess noise penalty of $3\times10^{-4}$ SNU.
  Discrete-modulation schemes, especially QPSK, have a performance penalty with respect to Gaussian modulation \cite{Lin2019, Denys2021explicitasymptotic}. Our results however still yield a significant key rate of 15 Mbit/s under \cite{Lin2019}. We hence have successfully demonstrated, for the first time, joint classical/QKD transmission where the classical channel provides quantum carrier phase and frequency recovery. We believe it is possible to improve on these results in terms of excess noise and thus distance by using higher resolution ADCs at Bob's. We leave this for future work, together with the use of higher order modulation formats for the quantum channel since they provide higher secure key rates. Leveraging coherent modulation formats and digital signal processing, our work hence opens new paths towards cost-effective integration of quantum and classical communications.

\section{Acknowledgements}
This project has received funding from the European Union’s Horizon 2020 research and
innovation programme under grant agreement CIVIQ No 820466. R. Aymeric acknowledges support from Télécom Paris Alumni. 
\bibliographystyle{osajnl.bst}
\bibliography{references}

\end{document}